\documentclass[twocolumn,showpacs,preprintnumbers,amsmath,amssymb]{revtex4}
\usepackage{graphicx}
\usepackage{dcolumn}
\usepackage{bm}
\newcommand{\bea}{\begin{eqnarray}}
\newcommand{\eea}{\end{eqnarray}}
\newcommand{\be}{\begin{equation}}
\newcommand{\ee}{\end{equation}}
\newcommand{\xbf}[1]{\mbox{\boldmath $ #1 $}}
\def\shiftdown#1{#1\llap{\lower.04ex\hbox{#1}}}

\begin{document}
\title{{\Large Structure of strange baryons}}
\thanks{Proceedings of IX International Conference 
on Hypernuclear and Strange Particle 
Physics, HYP 2006, Oct.~10-14, 2006, Mainz, Germany, Springer-Verlag 2007, 
pg. 329.}

\author{A. J. Buchmann}
\email{alfons.buchmann@uni-tuebingen.de}
\affiliation{Institute for Theoretical Physics \\
University of T\"ubingen \\
Auf der Morgenstelle 14 \\
D-72076 T\"ubingen, Germany}

\begin{abstract}
The charge radii and quadrupole moments of baryons with 
nonzero strangeness are calculated using a parametrization method
based on the symmetries of the strong interaction.
\end{abstract}

\pacs{14.20.Jn, 13.40.Gp, 11.30.Ly}
\maketitle
\section{Introduction}
\label{sec:introduction}

The discovery of the first strange meson by Rochester and Butler~\cite{Roc47} 
marked the beginning of a new era in subatomic physics. Since then
a great deal has been learned about the strong and electroweak interactions of 
mesons and baryons. However, our knowledge of their spatial structure 
is still rather limited. In contrast to the $N$ and $\Delta$, 
which have been investigated with increasing 
precision~\cite{jon00,Ber03,kel02}, 
little is known about the spatial structure of their strange siblings
with whom they form the octet and decuplet baryon families. 
For example, while proton and neutron
charge radii were already measured half a century ago~\cite{cha56}, 
a first determination of the 
$\Sigma^-$ charge radius has only recently become 
possible~\cite{Esc01}. Up until now, there have been no experiments 
that pertain to the shape of hyperons~\cite{Poc03}. 
In addition, with respect to theory we are still lacking 
a thorough understanding of the structure of most hadrons.
Various model calculations of hyperon charge radii~\cite{chargeradcalc} 
and quadrupole moments~\cite{quadmomcalc} differ considerably in their 
predictions. The reasons for these differences are often unclear. 
To gain a better understanding, 
we investigate to what extent these structural features 
are determined by the symmetries of the strong interaction. 

\section{Strong interaction symmetries}
\label{sec:symmetry}

Invariance of the strong interaction under SU(2) isospin
transformations leads to isospin conservation and the appearance
of degenerate hadron multiplets with fixed isospin, such as 
the pion triplet and nucleon doublet. 
In the 1950s numereous new meson and baryon isospin multiplets 
were discovered~\cite{Pai86}. 
The unusually long lifetime of these new particles 
(fast production but slow decay) was explained by Pais, Gell-Mann, and 
Nishijima~\cite{Gel53} by invoking a new symmetry principle and 
additive quantum number called `strangeness'. 
The latter was assumed to be 
conserved in strong and electromagnetic interactions (production) 
but violated in weak interactions (decay), thus explaining the 
long lifetimes of strange hadrons.

\begin{figure}
\begin{center}
\resizebox{0.45\textwidth}{!}{
\includegraphics{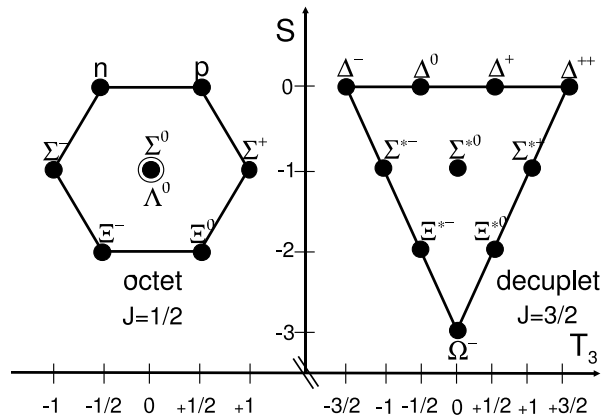}}
\end{center} 
\vspace{-0.5cm} 
\caption{\label{fig:su3}
SU(3) flavor octet and decuplet of ground-state baryons characterized by their 
strangeness $S$ (vertical axis) and isospin component $T_3$ 
(horizontal axis). }
\end{figure}

Further considerations led Gell-Mann to propose 
that strong interactions conserve not only isospin and strange\-ness,
but are also invariant under the higher SU(3) flavor 
symmetry~\cite{Gel64} which ties isospin multiplets with 
different isospin $T$ and strangeness $S$ to larger degenerate 
multiplets of particles with the same spin ${\bf J}$ and parity $P$, e.g., 
to octets and decuplets (see Fig.~\ref{fig:su3}). It was soon recognized
that these higher multiplets emerge because baryons are composed 
of the same spin 1/2 flavor triplet quarks, which are merely coupled to a 
different total spin and flavor. 
An even higher strong interaction symmetry is SU(6)
spin-flavor symmetry, which unites the spin 1/2 flavor octet baryons
($2 \times 8$ states), and the
spin 3/2 flavor decuplet baryons ($4 \times 10$ states) 
into a common {\bf 56}-dimensional mass degenerate
supermultiplet~\cite{Gur64,Beg64}. 
There are numerous successful predictions based on SU(6) spin-flavor
symmetry.
For example, while SU(3) flavor symmetry alone does not suffice to
uniquely determine the ratio of proton and neutron magnetic moments,
SU(6) spin-flavor symmetry~\cite{Beg64} leads to the prediction
$\mu_p/\mu_n=-3/2$, which is in excellent agreement with the experimental 
result -1.46. Another example is the G\"ursey-Radicati 
mass formula~\cite{Gur64} which explains why the Gell-Mann Okubo mass 
formula works so well for both octet and decuplet baryons with the same 
numerical coefficients. Without an underlying spin-flavor symmetry this would 
remain mysterious.

Thus, SU(6) is a good symmetry in baryon physics, and the question 
arises whether it is a symmetry of quantum chromodynamics. 
In an $1/N_c$ expansion, where $N_c$ denotes the number of 
colors, it has been shown that QCD possesses a spin-flavor symmetry which 
is exact in the large $N_c$ limit~\cite{sak84,das95}. For finite 
$N_c$, spin-flavor symmetry breaking operators can be classified
according to the powers of $1/N_c$ associated with them.
It turns out that higher orders of spin-flavor symmetry breaking
are suppressed by correspondingly higher powers of $1/N_c$.
This leads to a perturbative expansion scheme
for QCD processes that works at all energy scales and
provides a connection between broken SU(6) spin-flavor symmetry
and the underlying quark-gluon dynamics~\cite{Wit79,leb98}.

\section{Method} 

Alternatively, for $N_c=3$ we may use a straightforward model-independent 
parametrization method developed by Morpurgo~\cite{Mor89}, 
which incorporates SU(6) symmetry 
and its breaking similar to the $1/N_c$ expansion.
The basic idea is to {\it formally} define, for the observable at
hand, a QCD operator $\Omega$ and QCD eigenstates $\vert B \rangle$
expressed explicitly
in terms of quarks and gluons. The corresponding matrix element
can, with the help of the unitary operator $V$, be reduced to an
evaluation in the basis of auxiliary three-quark states
$\vert\Phi_B \rangle $
\begin{equation}
\label{map}
\left \langle B \vert \Omega \vert B \right \rangle =
\left \langle \Phi_B \vert
V^{\dagger}\Omega V \vert \Phi_B \right \rangle =
\left \langle W_B \vert
{ {\cal O}} \vert W_B \right \rangle \, .
\end{equation}
The auxiliary states $\vert \Phi_B \rangle$
are pure three-quark states with orbital angular momentum
$L=0$.  The spin-flavor wave functions~\cite{Lic78}
contained in $\vert \Phi_B \rangle$ are denoted by $\vert W_B\rangle $.
The operator $V$ dresses the pure three-quark
states with $q\bar q$ components and gluons and
thereby generates
the exact QCD eigenstates $\vert B \rangle $. Furthermore,
it is implied that $V$
contains a Foldy-Wouthuysen transformation allowing the
auxiliary states to be written in terms of Pauli spinors.

One then writes the most general expression for the operator
${ {\cal O}}$ that is
compatible with the space-time and inner QCD symmetries.
Generally, this is a sum of one-, two-, and three-quark
operators in spin-flavor space multiplied by {\it a priori} unknown constants
which parametrize the orbital and color space
matrix elements.
Empirically, a hierarchy in the importance
of one-, two-, and three-quark operators is found. 
This fact can be understood
in the $1/N_c$ expansion where
two- and three-quark operators describing second and third 
order SU(6) symmetry breaking
are usually suppressed by powers of $1/N_c$ and $1/N_c^2$ respectively,  
compared to one-quark operators associated with first order symmetry breaking.
The method has been used to calculate various properties of
baryons and mesons~\cite{Mor89,Mor99,Mor01,Hen02}.
In the next section, we apply it 
to baryon charge radii and quadrupole moments.

\section{Observables}
\label{sec:observables}

Information on baryon structure is contained 
in the charge monopole form factor $G_{C0}(q^2)$, where $q^2$ 
is the four-mo\-men\-tum transfer of the virtual photon. 
In the Breit frame, 
the Fourier transform of $G_{C0}(q^2)$ corresponds to
the charge density $\rho({r})$, describing the {\it radial} dependence 
of the bary\-on charge distribution. 
Its lowest radial moment is the baryon charge radius $r^2_B$. 

However, the charge density need not be spherically symmetric,
i.e., in general $\rho({\bf r}) \ne \rho(r)$. The geometric shape of a baryon 
is determined by its intrinsic quadrupole moment~\cite{Hen01}, 
which can be inferred from the observable spectroscopic quadrupole moment 
$Q_B$.
The latter corresponds to the charge quadrupole form factor $G_{C2}(q^2)$ 
at zero momentum transfer.
For spin 1/2 baryons, which do not have a spectroscopic quadrupole moments 
due to angular momentum selection rules, one may still obtain 
information on their intrinsic quadru\-pole
moments from measurements of electric $(E2)$ and Coulomb $(C2)$ quadru\-pole
transitions to excited states~\cite{Ber03}. 
   
The lowest moments of the charge density operator 
${\rho}$ are obtained from a multipole expansion at low momentum transfers. 
Up to $q^2$ contributions one has
\be
\rho(q) = 
e - \frac{q^2}{6} \, {r}^2
- \frac{q^2}{6} {\cal Q} + \cdots
\ee
The first two terms arise from the spherically symmetric monopole
part, while the third term is obtained from the quadrupole
part of $\rho$. They characterize the total charge ($e$),
spatial extension ($r^2$), and shape (${\cal Q}$) of the system.

\subsection{Charge radii}
\label{subsec:radii}

According to the method outlined in the previous section,
the charge radius operator can be expressed 
as a sum of one-, two-, and three-quark terms in spin-flavor space as
\begin{equation}
\label{para1}
{r}^2 =  A \sum_{i=1}^3 e_i {\bf 1} + 
B\sum_{i \ne j}^3 e_i \, \xbf{\sigma}_i \cdot \xbf{\sigma}_j  + 
C\!\! \sum_{i \ne j \ne k }^3 e_k \, \xbf{\sigma}_i \cdot \xbf{\sigma}_j, 
\end{equation}
where
$e_i=(1 + 3 \tau_{i \, z})/6$ and $\xbf{\sigma}_i$ are 
the charge and spin of the i-th quark.
Here, $\tau_{i \, z}$ denotes the $z$ component of the Pauli isospin matrix.
These are the only allowed spin scalars that can be constructed from the 
generators of the spin-flavor group~\cite{comment2}. 
The constants $A$, $B$, and $C$ 
parametrizing the orbital and color matrix elements 
are determined from experiment.

\begin{table}[t]
\begin{tabular}{ l | l } 
                &                     \\ 
\hline
$n$             & $\phantom{A}-2B+4C$      \\
$p$             & $ A -6C$   \\
$\Sigma^{-}$    & $[A(2+\zeta) + 2(B+C) (1-2\zeta-2\zeta^2)]/3 $   \\
$\Sigma^{0}$    & $[A(1-\zeta) + 
B(1-2\zeta+4\zeta^2) -2C(1+\zeta+\zeta^2)]/3$ \\
$\Lambda^{0}$   & $A(1-\zeta)/3 -B + 2C $ \\
$\Lambda\Sigma$ & $ \phantom{A +} 
-\sqrt{3} (B \zeta  -C(\zeta+\zeta^2))$  \\ 
$\Sigma^{+}$    & $[A(4-\zeta) + 4B(1 -2\zeta +\zeta^2) 
-2C(1 +4\zeta+4\zeta^2)]/3 $ \\
$\Xi^{-}$       & $[A(1+2\zeta) - 2(B+C)(2\zeta + 2\zeta^2 -\zeta^3)]/3$  \\
$\Xi^{0}$ &    $[A(2-2\zeta) -
2B(4\zeta-2\zeta^2+\zeta^3) +4C(\zeta+\zeta^2+\zeta^3)]/3$ \\
\hline
\hline
$\Delta^-$      & $ A +2B +2C$   \\
$\Delta^{0}$    & $0$   \\
$\Delta^{+}$    &  $ A +2B +2C $    \\
$\Delta^{++}$   & $ A +2B +2C $   \\
$\Sigma^{\ast -}$ & $[A(2+\zeta) + 2(B+C)(1 +\zeta +\zeta^2)]/3$ \\
$\Sigma^{\ast 0}$ & $[A(1-\zeta) + 
B(1 +\zeta -2\zeta^2) -C(2 -\zeta-\zeta^2)]/3 $ \\
$\Sigma^{\ast +}$ & $[A(4-\zeta)+ 2B(2 +2\zeta -\zeta^2) 
-2C(1 -2\zeta-2\zeta^2)]/3 $ \\
$\Xi^{\ast -}$   & $[A(1+2\zeta) + 2(B+C)(\zeta + \zeta^2 +\zeta^3)]/3$  \\
$\Xi^{\ast 0}$ & $[A(2-2\zeta) + 
2B(2\zeta-\zeta^2-\zeta^3) -2C(\zeta+\zeta^2-2\zeta^3)]/3$\\
$\Omega^{-}$ & $ A\zeta +2(B+C)\zeta^3$ \\
\hline
\end{tabular}
\vspace{0.2cm} 
\caption[chargeradii]{\label{tab:baryonradii} Baryon charge radii  
denoted by the particle symbols with one-quark $(A)$, two-quark ($B$), 
and three-quark ($C$) contributions. Flavor symmetry breaking 
is characterized by the ratio of u-quark and s-quark masses $\zeta=m_u/m_s$.
The flavor symmetry limit is obtained for $\zeta=1$.}
\end{table}

To estimate the degree of SU(3) flavor symmetry breaking
we insert in Eq.(\ref{para1}) a linear and cubic quark mass dependence as
\be
\label{su3break} 
{\bf 1} \, \rightarrow {\bf 1} \, \, (m_u/m_s),  \qquad 
\xbf{\sigma}_{i} \cdot \xbf{\sigma}_{j}  \rightarrow  \xbf{\sigma}_{i} \cdot
\xbf{\sigma}_{j} \, \, m_u^3/(m_i^2 m_j). 
\ee
These replacements are motivated by the different charge radii of up- and 
strange quarks and the flavor depen\-dence 
of the gluon exchange current diagram~\cite{Hen02}.  
Flavor symmetry breaking is then characterized by the ratio
$\zeta=m_u/m_s$ of $u$ and $s$ quark masses. The latter is  
determined from the magnetic moment of the $\Lambda$ hyperon.
Our treatment of flavor symmetry breaking is not exact. 
Improvements are possible by introducing 
additional operators and constants
in Eq.(\ref{para1}). However, there would then be so 
many undetermined constants
that the theory can no longer make predictions. We expect that our 
approximate treatment of flavor symmetry breaking captures the most important 
physical effects. 
We use the same mass for $u$ and $d$ quarks
to preserve the SU(2) isospin symmetry of the strong interaction
that is known to hold to a very good accuracy.
\begin{table}[htb]
\begin{center}
\begin{tabular}{ l|  r | r | r }
 &   $(a)$ &  $(b)$ & $(c)$ \\
 \hline
$n$    &    -0.116       & -0.116 & -0.116       \\
$p$     &    0.779       &  0.779  & 0.779     \\
$\Sigma^{-}$  & 0.610        & 0.641 & 0.672    \\
$\Sigma^{0}$  & 0.122        & 0.125  & 0.128 \\
$\Lambda^{0}$  & 0.035        & 0.043 & 0.050   \\
$\Lambda^{0}\Sigma^0$ & -0.058 &  -0.062 & -0.066     \\
$\Sigma^{+}$  & 0.855        &  0.891  & 0.928 \\
$\Xi^{-}$     & 0.501        &  0.510 & 0.520  \\
$\Xi^{0}$     & 0.120        &0.126  & 0.132 \\
\hline 
\hline 
$\Delta^{-}$    & 0.783           & 0.895  & 1.011       \\
$\Delta^{0}$     & 0          &      0    & 0\\
$\Delta^{+}$     & 0.783          & 0.895 & 1.011     \\
$\Delta^{++}$     & 0.783         & 0.895  & 1.011 \\
$\Sigma^{\ast -}$ &  0.669        &  0.756 & 0.845  \\
$\Sigma^{\ast 0}$  & 0.108       &  0.117 & 0.127 \\
$\Sigma^{\ast +}$  & 0.869        & 0.988 & 1.086    \\
$\Xi^{\ast -}$     & 0.561        & 0.625 & 0.692   \\
$\Xi^{\ast 0}$    &  0.206        & 0.225 & 0.244 \\
$\Omega^-$        & 0.457         & 0.504 & 0.553 \\
\hline
\end{tabular}
\vspace{0.2cm} 
\caption[charadnum]{\label{tab:baryonradiinum}
Numerical charge radii in [fm$^2$]
according to Table~\ref{tab:baryonradii}. 
We use the measured charge radii 
$r^2_n=-0.1161(22)$ fm$^2$~\cite{Eid04}, 
$r^2_p=0.779(25)$ fm$^2$~\cite{Ros00}, and $\zeta=0.613$ as input.
Three different values for $r^2_{\Sigma^-}$ are used as input:
$r^2_{\Sigma^-}=0.61(21)$ fm$^2$~\cite{Esc01} (a), 
$r^2_{\Sigma^-}=0.64$ fm$^2$ (b),
and $r^2_{\Sigma^-}=0.67$ fm$^2$ (c).}
\end{center}
\end{table}

Baryon charge radii are then calculated by evaluating 
matrix elements of the operator in Eq.(\ref{para1}) 
including the substitutions in Eq.(\ref{su3break}) between three-quark 
spin-flavor wave functions $\vert W_B \rangle$
\be
r_B^2 = \langle W_B \vert {r}^2 \vert W_B \rangle .
\ee
For charged baryons, $r^2_B$
is normalized by dividing by the baryon charge. 
The results for octet and decuplet baryons are summarized in 
Table~\ref{tab:baryonradii}. The two- and three-quark results
agree with those in Ref.~\cite{Buc03} for $N_c=3$ 
after obvious redefinition of the constants.

From Table~\ref{tab:baryonradii} one readily observes 
that in the SU(6) symmetry limit all ground state bary\-ons 
have the same charge radius $r^2_B=A\, e $, where $e$ is the baryon charge.
In particular, the charge radii of the neutral baryons are zero.
Inclusion of the spin-dependent two- and three-quark operators break 
SU(6) spin-flavor symmetry and split
the charge radii of octet and decuplet baryons by decreasing the former
and increasing the latter, e.g.
\be
r^2_{\Sigma^{*\, -}} - r^2_{\Sigma^-} 
=  r^2_{\Xi^{*\, -}} - r^2_{\Xi^-}
= 2 (B+C) (\zeta + \zeta^2) > 0.  
\ee 
Another consequence is that 
the neutral octet charge radii are definitely nonzero.
If third order SU(6) symmetry breaking, i.e. the $C$ term,  
can be neglected,
the following relation holds~\cite{Mor99,Buc97}
\be
r^2_{\Delta^+}  -  r^2_{p}  =  -r^2_{n}, 
\ee
which shows that the octet-decuplet charge radius splitting is of the same 
order as the neutron charge radius.

Because there are 19 experimental charge radii (8 diagonal octet, 
1 transition octet, and 10 diagonal decuplet) and 3 constants, 
Table~\ref{tab:baryonradii} contains 16 relations among the baryon 
charge radii, some of which are independent of the SU(3) symmetry breaking
parameter $\zeta$ and thus hold irrespective of how
badly SU(3) flavor symmetry is broken. An example of the latter type 
is the $\Sigma$ equal spacing rule, 
\be
\label{sigmaequalspacingrule}
r^2_{\Sigma^+} - r^2_{\Sigma^-} = 2 \, r^2_{\Sigma^0}, \qquad
r^2_{\Sigma^{\ast +}} - r^2_{\Sigma^{\ast -}} = 2 \, r^2_{\Sigma^{\ast 0}},
\ee 
which is already a consequence of the assumed isospin symmetry.
Another example is the equal spacing rule for decuplet charge radii
\be
\label{equalspacingrule}
3\, (r^2_{\Xi^{\ast -}} - r^2_{\Sigma^{\ast -}}) = r^2_{\Omega^{\ast -}}
-r^2_{\Delta^{\ast -}},
\ee
which the reader may easily verify using 
Table~\ref{tab:baryonradii}. Analogous relations 
hold for baryon quadrupole moments~\cite{Hen02}. 

To make numerical predictions, we express the three
parameters $A$, $B$, and $C$ in Eq.(\ref{para1}) in terms of the three 
measured charge radii
$r^2_p$,  $r^2_n$, and $r^2_{\Sigma^-}$ using the results in 
Table~\ref{tab:baryonradii} 
\bea
\label{gpconstants}
A(3-\zeta -2\zeta^2)\! \!& = & \! \!(1-2\zeta -2\zeta^2) (r_p^2 -r_n^2) 
+ 3r_{\Sigma^-}^2,  \nonumber \\
-3B(3-\zeta -2\zeta^2) \! \! & = & \! \! (2 +\zeta)r_p^2 
+ \frac{1}{2}( 7+ \zeta -2\zeta^2)r_n^2 
- 3\, r_{\Sigma^-}^2,  \nonumber \\ 
-6C(3-\zeta -2\zeta^2) \! \! & = & \! \!(2 +\zeta)r_p^2 
- \frac{1}{2}( 2 -4\zeta - 4\zeta^2)r_n^2 
- 3r_{\Sigma^-}^2. \nonumber
\eea
With the experimental radii and $\zeta=0.613$ we obtain 
$A=0.723$ fm$^2$, $B=0.039$ fm$^2$, $C=-0.009$ fm$^2$ 
and the numerical results 
shown in Table~\ref{tab:baryonradiinum} (a). 
For comparison, we also use 
$r^2_{\Sigma^-}=0.64 (0.67)$ fm$^2$ as input, which  
leads to $A=0.779 (0.873)$ fm$^2$, $B=0.058 (0.077)$ fm$^2$, 
$C=0 (0.010)$ fm$^2$, and the results labelled (b) and (c) in 
Table~\ref{tab:baryonradiinum}. 
Thus, while the constant $B>0$ is fixed by the negative neutron
charge radius, the sign of $C$ cannot be reliably determined.
In any case, the $C$ term is definitely smaller 
than the $B$ term by at least a factor of $1/N_c$. 
\begin{table}[b]
\begin{tabular}{l | l}
\hline
$n\to \Delta^0$  &  $2\sqrt{2}(B'-2C')$       \\
$p\to \Delta^+$   & $2\sqrt{2}(B'-2C')$  \\
$\Sigma^- \to \Sigma^{\ast -}$ & $-\sqrt{2} \, (2B'+2C')\, 
(2-\zeta-\zeta^2)/3$    \\
$\Sigma^0 \to \Sigma^{\ast 0}$ & 
$\sqrt{2} [2B'(2-\zeta+2\zeta^2)-2C'(4 + \zeta +\zeta^2)]/6 $ \\
$\Lambda^0 \to \Sigma^{\ast 0}$ & $\sqrt{6}[2B' 
\zeta - 2C'(\zeta + \zeta^2)]/2 $  \\
$\Sigma^+ \to \Sigma^{\ast +}$ & $2\sqrt{2}\, \lbrack B' \,(4-2\zeta+\zeta^2) 
- 2C' \,(1+\zeta +\zeta^2) \rbrack /3 $ \\
$\Xi^- \to \Xi^{\ast -}$ & $-\sqrt{2} \,(2B'+2C')\, 
(\zeta+\zeta^2-2\zeta^3)/3$ \\
$\Xi^0 \to \Xi^{\ast 0}$ & 
$\sqrt{2}[2B'(2\zeta -\zeta^2 + 2\zeta^3)
-2C'(\zeta + \zeta^2 + 4\zeta^3)]/3$ \\
\hline
\hline
$\Delta^{-}$      & $ -4B' -4C'$    \\
$\Delta^{0}$      &     0         \\
$\Delta^{+}$      & $4B' +  4C'$     \\
$\Delta^{++}$     & $8B' + 8C'$     \\
$\Sigma^{\ast -}$ & $-(4B'+4C') (1+\zeta+\zeta^2)/3 $      \\
$\Sigma^{\ast 0}$ & $ [2B' (1+\zeta-2\zeta^2) - 2C'(2-\zeta-\zeta^2)]/3$   \\
$\Sigma^{\ast +}$ & $ [4B'(2 + 2\zeta -\zeta^2) -4C'(1-2\zeta-2\zeta^2)]/3$  \\
$\Xi^{\ast -}$    &  $-(4B'+4C')(\zeta + \zeta^2 +\zeta^3)/3$     \\
$\Xi^{\ast 0}$    & $[4B'(2\zeta-\zeta^2-\zeta^3) 
-4C'(\zeta+\zeta^2-2\zeta^3)]/3$ \\ 
$\Omega^-$        & $-(4B' + 4C')\zeta^3 $          \\
\hline
\end{tabular}
\caption[C2 moments]{\label{tab:quadmo}
Transition and diagonal baryon quadrupole moments
denoted by the particle symbols with two-quark (B') and three-quark (C')
contributions. From Ref.~\cite{Hen02}.}
\end{table}
\begin{table}[b]
\begin{center}
\begin{tabular}{l | r | r | r}
 &   $(a)$ &  $(b)$ & $(c)$ \\
\hline
$n\to \Delta^0$  &  $-0.082$   & -0.082 & -0.082    \\
$p\to \Delta^+$   & $-0.082$  & -0.082 & -0.082 \\
$\Sigma^- \to \Sigma^{\ast -}$ & $0.014$ & 0.028 & 0.042   \\
$\Sigma^0 \to \Sigma^{\ast 0}$ &  $-0.031$ & -0.029 & -0.028 \\
$\Lambda^0 \to \Sigma^{\ast 0}$ & $-0.041$ &-0.044 & -0.046  \\
$\Sigma^+ \to \Sigma^{\ast +}$ & $-0.076$ & -0.086 & -0.097\\
$\Xi^- \to \Xi^{\ast -}$ & $0.007$ & 0.014 &  0.022\\
$\Xi^0 \to \Xi^{\ast 0}$ & $ -0.033 $ & -0.036 & -0.039\\
\hline
\hline
$\Delta^{-}$      & $ 0.060$  &  0.116 &   0.174 \\
$\Delta^{0}$      &     $0$  & 0  &    0 \\
$\Delta^{+}$      & $-0.060$  & -0.116  &    -0.174\\
$\Delta^{++}$     & $-0.120$  & -0.232  &  0.348 \\
$\Sigma^{\ast -}$ & $ 0.039$ & 0.077 &   0.115   \\
$\Sigma^{\ast 0}$ & $ 0.014$ & -0.017 &  -0.019 \\
$\Sigma^{\ast +}$ & $-0.069$ & -0.110& -0.153  \\
$\Xi^{\ast -}$    &  $0.024$ & 0.047 &  0.071   \\
$\Xi^{\ast 0}$    & $-0.019$ & -0.023 & -0.029\\ 
$\Omega^-$        & $0.014$  & 0.027 &  0.040      \\
\hline
\end{tabular}
\caption[C2 momentsnum]
{\label{tab:quadmonum}
Numerical quadrupole moments in [fm$^2$] according 
to Table~\ref{tab:quadmo}
using the parameter sets (a)-(c) of sect.~\ref{subsec:radii}.}
\end{center}
\end{table}

Table~\ref{tab:baryonradiinum} shows that octet baryon charge radii 
are ordered  as $r^2_{\Sigma^+}\! > 
\!r^2_{p} \!> \!r^2_{\Sigma^-} \!> \!r^2_{\Xi^-}$.
For decuplet baryons one finds 
$r^2_{\Sigma^{\ast +}} \!> \!r^2_{\Delta^+}\! > \!r^2_{\Sigma^{\ast -}}
\! > \!r^2_{\Xi^{\ast -}} \!> \! r^2_{\Omega^-}$. In both flavor multiplets
the decrease in each step is of order $\vert r^2_n\vert$.
Note that only $r^2_n$ and $r^2_{\Lambda \Sigma}$ 
are negative~\cite{comment5}.  With respect to the size of  
SU(6) symmetry breaking, we obtain for the parameter sets (a)-(c)
a relative charge radius splitting 
$(r^2_{\Delta^+} -r^2_{p})/[(r^2_{\Delta^+} +r^2_{p})/2]$ of $0.5\%$, $14\%$,
and $26\%$. For $r^2_{\Sigma^-}=0.91(36)$ fm$^2$~\cite{Esc01} 
(WA89 experiment) the splitting would 
be 83$\%$, which is too large, if the relative octet-decuplet splitting
for charge radii and masses is similar, as we expect.
 
\subsection{Quadrupole moments}
\label{subsec:quadmom}
The charge quadrupole operator is composed of a two- and three-body term
in spin-flavor space
\bea
\label{para2}
{\cal Q}  & = & B'\sum_{i \ne j}^3 e_i 
\left ( 3 \sigma_{i \, z} \sigma_{ j \, z} -
\xbf{\sigma}_i \cdot \xbf{\sigma}_j \right ) \nonumber \\
& + & C'\!\!\sum_{i \ne j \ne k }^3 e_k 
\left ( 3 \sigma_{i \, z} \sigma_{ j \, z} -
\xbf{\sigma}_i \cdot \xbf{\sigma}_j \right ). 
\eea
Baryon decuplet quadrupole moments $Q_{B^*}$ and octet-decu\-plet tran\-sition
quadrupole moments $Q_{B \to B^*}$ are obtained by calculating the
matrix elements of the quadrupole operator
in Eq.(\ref{para2}) between the
three-quark spin-flavor wave functions $\vert W_B \rangle $
\begin{eqnarray}
\label{matrixelements}
Q_{B^*} & = &\left \langle W_{B^*} \vert {\cal Q}
\vert W_{B^*} \right \rangle , \nonumber \\
Q_{B \to B^*}
& = & \left \langle W_{B^*} \vert {\cal Q} \vert W_B \right \rangle,
\end{eqnarray}
where $B$  denotes a spin 1/2 octet baryon and $B^*$ a member of the
spin 3/2 baryon decuplet.  The ensuing quadrupole moment relations
have been discussed earlier~\cite{Hen02}. 
Table~\ref{tab:quadmo} reproduces the main results.

Interestingly, SU(6) symmetry does not only lead to 
charge radius and quadrupole moment relations, 
but also furnishes relations between
both sets of observables~\cite{Buc06,Buc04}. 
This can be seen from a comparison
of Table~\ref{tab:baryonradii} and Table~\ref{tab:quadmo} using the
relations $B'=-B/2$ and $C'=-C/2$~\cite{comment6}.  
For example, we find that the
$N \to \Delta$ quadrupole moment is related to the neutron charge 
radius as~\cite{Buc97}
\be 
Q_{p \to \Delta^+} = Q_{n \to \Delta^0} = \frac{1}{\sqrt{2}}\, r_n^2,
\ee 
which is experimentally well satisfied~\cite{Tia03}. 
This relation also holds for the corresponding form factors 
up to momentum transfers in the GeV region~\cite{Buc04}.
In the SU(3) symmetry limit, the transition quadrupole moments
of nonnegative hyperons are proportional to $r^2_n$,
as can be verified from Tables~\ref{tab:baryonradii} and 
~\ref{tab:quadmo} for $\zeta=1$. Thus, the
neutron charge radius plays an important role. It sets the scale 
not only for the charge radius splitting within and between 
flavor multiplets but also for the size of quadrupole moments and the  
corresponding intrinsic baryon deformation~\cite{Hen01,Buc06,Buc04}. 
Another example of the predictive power of SU(6) spin-flavor symmetry 
is the $\Omega^-$ quadrupole moment 
\be 
Q_{\Omega^-} = \frac{1}{3-\zeta-2\zeta^2}\, \left ( 
3\, r^2_{\Sigma^-} -(2+\zeta) \,
(r^2_p + r^2_n) \right ) \zeta^3,
\ee
which has been expressed here in terms of the three measured 
baryon charge radii.
Finally, Table~\ref{tab:quadmonum} provides numerical predictions for 
baryon quadrupole moments using parameter sets (a)-(c) 
of sect.~\ref{subsec:radii}.

\section{Summary}
\label{sec:summary}
We have calculated baryon charge radii and quadrupole moments 
using a para\-metri\-zation method based on the symmetries
of QCD, and derived a number of charge radius and quadrupole moment 
relations. In addition, SU(6) symmetry leads to interesting
relations between these two sets of observables. 
Using the three measured radii as input, we have 
obtained numerical predictions for the remaining charge radii and 
baryon quadrupole moments.
Our results suggest that one can
obtain information on the shape of baryons both from quadrupole
moment and charge radius measurements.


\begin{thebibliography}{}
\bibitem{comment5}
In contrast to Ref.~\cite{Buc03}
we predict here $r^2_{\Lambda} > 0$. This is due to our inclusion of 
SU(3) flavor symmetry breaking in the one-quark term in Eq.(\ref{para1}).
\bibitem{comment6} In SU(6), the spin scalar 
charge radius and spin tensor quadrupole moment 
operators are related~\cite{Buc06}.
\end{thebibliography}
\end{document}